\documentclass[aps,epsf,twocolumn]{revtex4}
\usepackage{graphicx}
\usepackage{amsmath}
\usepackage{amssymb}
\usepackage{float}
\usepackage{caption}
\usepackage{subcaption}
\usepackage{booktabs}
\usepackage{color}
\usepackage{enumitem}
\usepackage{bm}

\usepackage[colorlinks=true,
            linkcolor=blue,
            urlcolor=blue,
            citecolor=blue]{hyperref}

\hypersetup{citecolor=blue}

\graphicspath{{./figures/}}

\begin{document}

\title{Slow driving induced multistability and remote synchronization in chaotic Chua's circuit}

\author{Tuhin Mahanty}
\affiliation{Department of Physics, Institute of Science, Banaras Hindu University, Varanasi UP 221005, India}

\author{Ayushi Saxena}
\affiliation{Department of Physics, Institute of Science, Banaras Hindu University, Varanasi UP 221005, India}

\author{Sangeeta Rani Ujjwal}
\affiliation{Department of Physics, Institute of Science, Banaras Hindu University, Varanasi UP 221005, India}


\begin{abstract}
We study the response of Chua's circuit driven by a chaotic signal of variable time-scale. We observe that when the frequency of the drive is significantly lower than that of the response and the driving strength is above a threshold, the Chua's circuit exhibits multiple stable attractors. The features of the attractors change as the driving strength $\varepsilon$ increases, for instance the attractors are double-scroll at low $\varepsilon$ and are single-scroll when $\varepsilon$ is high. We also investigate generalized synchronization (GS) between the drive and the response systems by employing the auxiliary system approach. When the drive is much slower than the response, we observe different scenarios of \emph {remote} synchronization (RS) between response and auxiliary units. In addition to complete synchrony between response and auxiliary systems indicating GS between drive and response, we notice that the response and auxiliary units can be lag synchronized and can also have correlated trajectories indicating novel forms of RS. The slow drive can induce multistability between these RS states which disappears as the frequency of drive increases and become equivalent to the response Chua's ciruit.     

\end{abstract}

\maketitle

\section{Introduction}

Study of chaotic synchronization ~\cite{syncbook, ottbook, boccaletti2002} has drawn considerable attention due to its potential applications in various areas such as chaos control~\cite{bowong2004, padmanaban2012} and secure communication~\cite{kocarev1995, kinzel2010}. Chaotic synchronization has been reported in several settings from mutually coupled systems~\cite{afraimovich1986} to systems in drive-response configuration~\cite{pecora1990, murali1994} and a wide variety of dynamical patterns from global synchrony, lag synchronization to generalized synchronization (GS) have been observed. Although a lot of investigation has been done to understand the dynamics of drive-response systems in different settings, most of these studies are limited to the case when drive and response have comparable time scales. There are some experimental studies that examine the synchronization of drive and response with time-scale mismatch~\cite{rulkov1996, kittel1998}. However, several aspects related to the emergent phenomena in a system subjected to a driving signal with variable frequency are still unexplored. Through this work, we try to explore this direction further.

When nonidentical systems are coupled in the drive-response arrangement, they can synchronize in the generalized sense, namely the state variable(s) of response can be expressed as a function of the state variable(s) of the drive~\cite{rulkov1995}. One method to analyse this generalized synchronization (GS) between drive and response is to use the auxiliary systems approach~\cite{abarbanel1996} wherein an exact copy of the response system called the auxiliary system is considered and if the response and the auxiliary units are completely synchronized then the response is said to be in generalized synchrony with the drive. This collective system of drive, response and auxiliary units can be viewed as a three-node network where the driving unit is acting like a hub and is unidirectionally coupled to the two unconnected systems (drive and auxiliary). An important feature observed in such networks with hubs is that of \emph{remote} synchronization (RS) where the nodes that are not directly connected to each other get synchronized whereas these nodes remain out of synchrony with the hub~\cite{bergner2012}. Experimentally electronic circuits are widely used to study chaotic synchronization and among these circuits Chua's circuit  ~\cite{chua1992-1,chua1994} has emerged as a popular choice since it is one of the simplest electronic circuits to exhibit chaotic dynamics~\cite{zhong1985}. Chua's circuit and its variants~\cite{murali1994-2} with different coupling schemes, for instance with resistive~\cite{chua1992-2} and capacitive coupling~\cite{liu2019} have been examined and various dynamical states are reported earlier~\cite{murali1992, itoh1994}.

In the present work, we study Chua's circuit in the chaotic regime driven by a chaotic signal of variable time-scale. We observe that when the time-scale of the drive is much slower than the response and the driving strength is above a threshold, the Chua's circuit exhibits bistability with the existence of two stable single-scroll attractors, though the dynamics of the uncoupled Chua's oscillator is on a double-scroll chaotic attractor. We also investigate the possibility of GS between the chaotic drive and Chua's circuit using the auxiliary system approach and found that when the driving strength is above a certain value the response and the auxiliary units are in complete synchrony implying GS between the drive and response systems. We note that the mismatch between the time scales of drive and response can induce novel types of RS wherein the response and auxiliary units not coupled directly oscillate with the same average frequency, however their dynamical variables are related to each other in different ways: the response and auxiliary units can be in complete synchrony, lag synchrony, and can have correlated trajectories. The slow varying drive can also induce multistability between these RS states. We analyse this effect of varying time-scale of the drive on the dynamics of the Chua's circuit and on the GS between drive and response both using circuit simulations on the platform Multisim and numerically by solving the dynamical equations. The response of Chua's circuit is studied by computing Lyapunov exponents of the system and it is observed that for a slow drive the largest conditional Lyapunov exponent (CLE) of the response is close to zero in a significant coupling range. The dynamics of the response system in this region is investigated using measures such as 0-1 test, Fourier transform and recurrence plots which indicate the transition of the dynamics from periodic to chaotic through intermittency. The drive is taken from a common base Colpitts oscillator ~\cite{roy2003}, however we get qualitatively similar results for other chaotic drives as well. 

The paper is arranged as follows: Sec.~\ref{sec:mod-des} describes the model considered for circuit simulation and numerical investigation. Multistability induced by driving is discussed in Sec.~\ref{sec:multy}. In Sec.~\ref{sec:gs} we analyse the generalized synchrony between drive and response by using the auxiliary system approach and present different scenarios of RS between response and auxiliary units. The investigation of the region of zero largest CLE using various tools are presented in Sec.~\ref{sec:inter}. The results are summarized in Sec.~\ref{sec:summary}.

\section{Model description} \label{sec:mod-des}

\begin{figure}
\centering
\includegraphics [scale=0.25, angle=0]{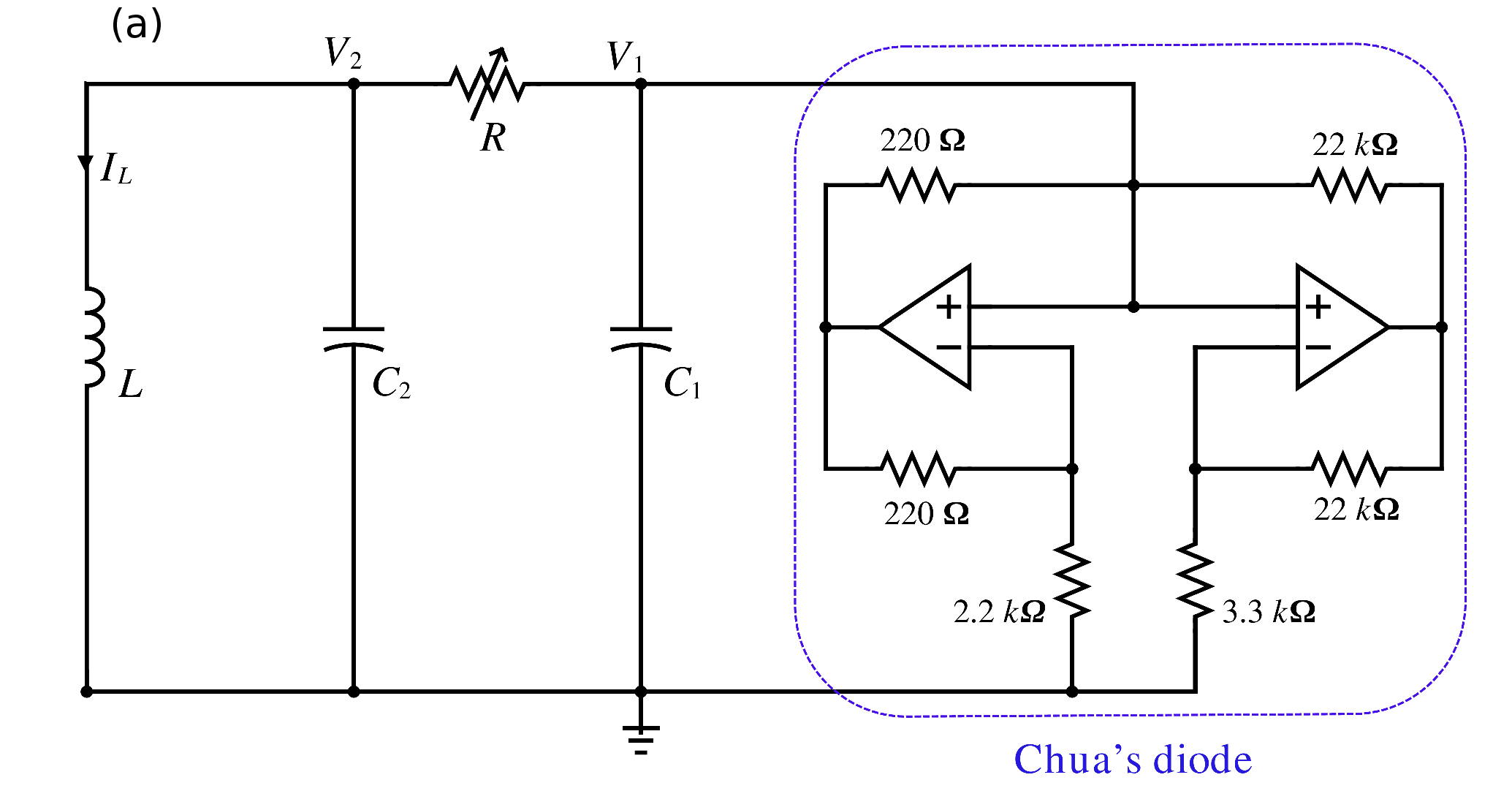} 
\includegraphics[width=0.40\textwidth]{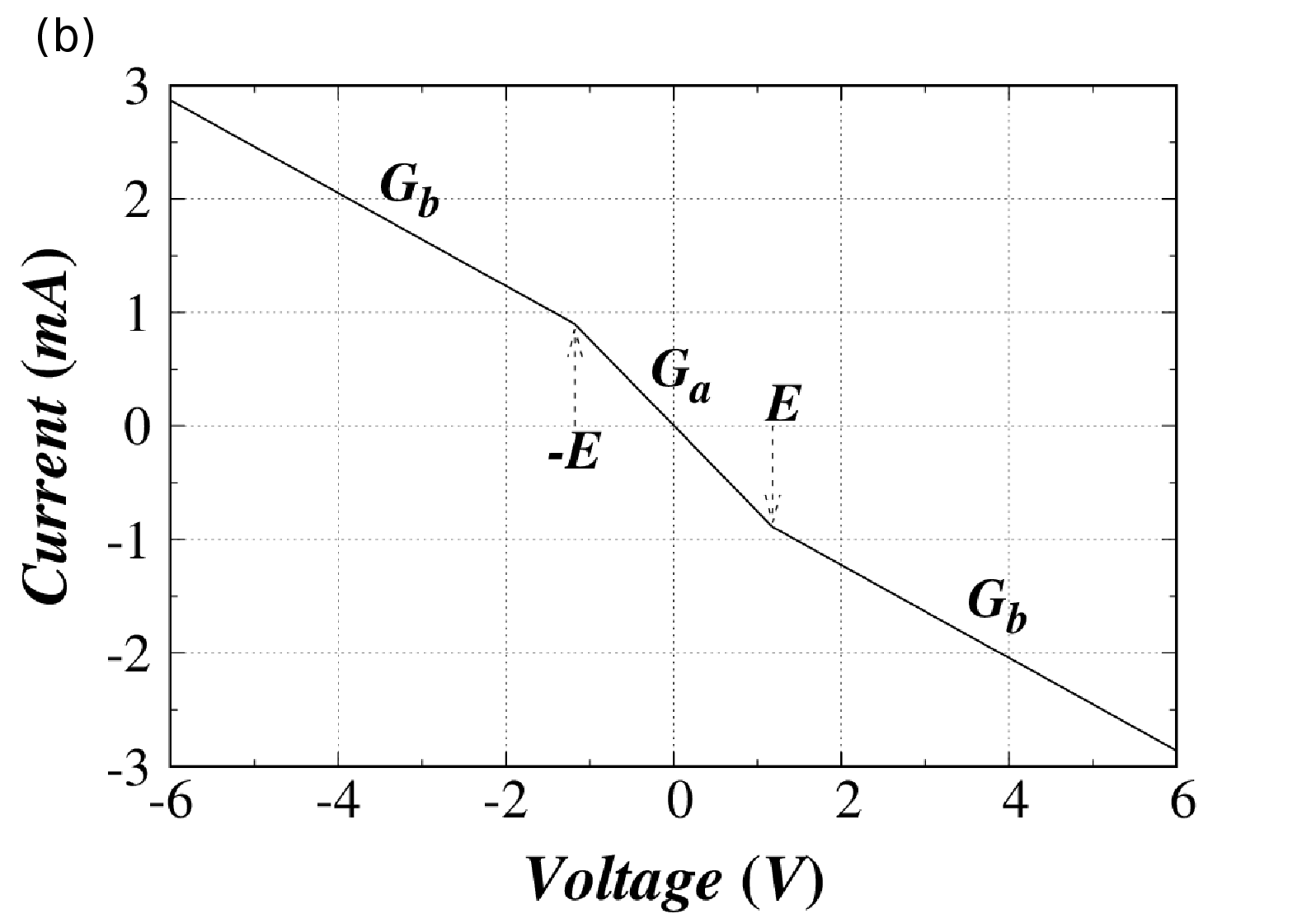}
\caption{(a)Circuit diagram of Chua's oscillator and (b)V-I characteristics of Chua's diode.}
\label{fig:Fig1}
\end{figure}

We consider Chua's circuit~\cite{chua1992-1,cruz1993} as the response system. The circuit representation of the Chua's oscillator is shown in Fig.~\ref{fig:Fig1}(a). The circuit parameters are taken to be ${C_1}=10 nF$, ${C_2}=100 nF$, $L=18 mH$ and  $R$ is the variable resistor that controls the dynamics of the Chua's oscillator. By applying Kirchhoff's circuital law to various branches of this circuit we get the following set of dynamical equations.

\begin{eqnarray}\label{eq:chua}
    \dot{V_1} &=& \frac{1}{RC_1}{[V_2-V_1-R f(V_1)]}, \nonumber \\   
     \dot{V_2} &=& \frac{1}{RC_2}(V_1-V_2+R I_L), \nonumber\\
     \dot{I_L} &=& -\frac{V_2}{L},
\end{eqnarray}
where the over-dot represents derivative with respect to time. The characteristic equation of the Chua's diode is given by  

 \begin{equation}
 f(V_1)=G_bV_1+\frac{1}{2}(G_a-G_b)(|{V_1+E}|-|{V_1-E}|), 
\end{equation}    
where $E$ is the turning point voltage and $G_a$, $G_b$ are the slopes obtained from the V-I characteristics of Chua's diode plotted in Fig.~\ref{fig:Fig1}(b). From the V-I characteristics of the Chua's diode we note that $G_a=-0.0007574$ $A/V$ , $G_b=-0.0004089$ $A/V$ and $|E|=1.18 V$. We use software Multisim to simulate the circuit in the laboratory environment. The asymptotic dynamics of the system is verified using numerical simulation of the governing dynamical equations (Eq.~\eqref{eq:chua}) using the Runge-Kutta 4th order method and the equivalence between the circuit and numerical simulations is established.

\begin{figure}
    \centering
    \includegraphics[scale=0.79, angle=0]{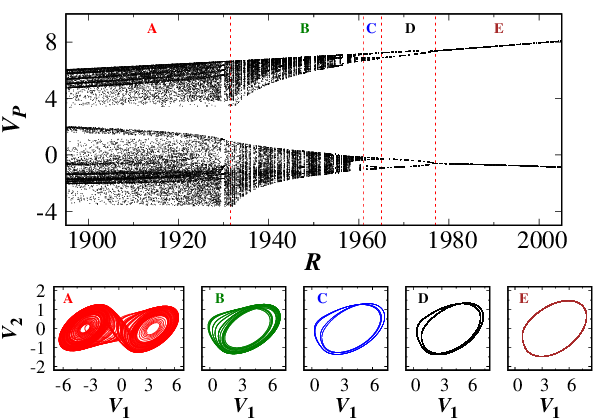}
    \caption{Upper panel shows the bifurcation diagram of Chua's oscillator obtained by plotting the maxima, $V_P$ in the time-series of $V_1$ as $R$ varies showing period-doubling route to chaos and regions of existence of different types of attractors: double-scroll chaotic (A), single-scroll chaotic (B), period-4 (C), period-2 (D), and period-1 (E). The attractors in different regions are shown in the lower panel at $R$ = 1920 $\Omega$ (A), 1950 $\Omega$ (B), 1963.5 $\Omega$ (C), 1970 $\Omega$ (D), and 2000 $\Omega$ (E) for a fixed initial condition~\cite{icfig3}. Please note that we have presented only one variant of the single-scroll attractors in the bistable region(B to E).}
    \label{fig:Fig2}
\end{figure}

An interesting feature of Chua's circuit equations (Eqs.~\eqref{eq:chua}) is that it remain invariant under parity transformation (($V_1, V_2, I_L) --> (-V_1, -V_2, -I_L)$) which can give rise to bistability of single-scroll attractors. In the Chua's circuit, as the resistance $R$ varies keeping other circuit elements constant, the dynamics of the Chua's circuit changes from periodic for larger $R$ to chaotic for smaller $R$ values ~\cite{tlelo2007, lv2021}. The bifurcation diagram depicting the transition from periodic to chaotic dynamics is plotted in Fig.~\ref{fig:Fig2} which shows a period-doubling route leading to chaotic dynamics as $R$ decreases. In Fig.~\ref{fig:Fig2} the maxima of the time-series of $V_1$ denoted by $V_P$ is plotted taking random initial conditions~\cite{sim-details} at each $R$. We note that depending upon the initial conditions, two variants of the single-scroll attractors are possible: one having positive $V_1$ values and the other having negative $V_1$ values in the phase space. In the lower panel of Fig.~\ref{fig:Fig2}, the attractors ($V_1$ vs $V_2$) obtained for different values of $R$ are shown. We have shown only one variant of the single-scroll attractor (having positive $V_1$ values) in the bistable region. As can be seen periodic attractors are obtained for high values of $R$: period-1 for $R$ = 2000 $\Omega$, period-2 for $R$ = 1970 $\Omega$, and period-4 for $R$ = 1963.5 $\Omega$. For lower values of $R$, for instance when $R$ = 1950 $\Omega$ single-scroll and $R$ = 1920 $\Omega$ double-scroll chaotic attractors are obtained.

\section{Multistability in the driven Chua's circuit}  \label{sec:multy}

\begin{figure}
    \centering
    \includegraphics[width=0.50\textwidth]{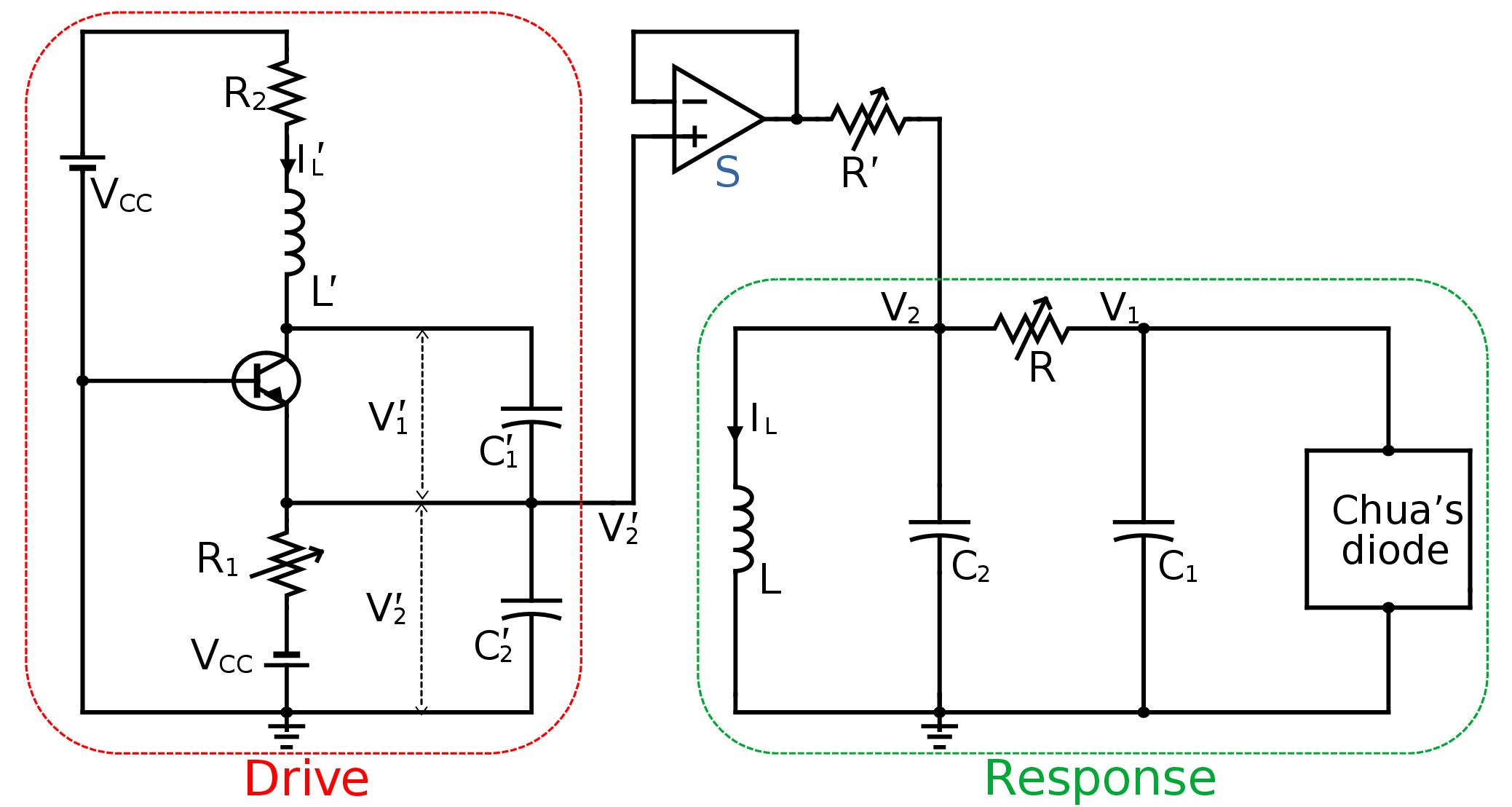}
    \caption{Circuit diagram of Chua's circuit driven by the common base Colpitts oscilator.}
    \label{fig:Fig3}    
\end{figure}

The Chua's circuit is driven by a chaotic drive taken from a common base Colpitts oscillator ~\cite{roy2003,kennedy1994}.
The circuit diagram of the Chua's circuit driven by Colpitts oscillator is shown in Fig.~\ref{fig:Fig3}. 
From the circuit the dynamical equations of the Colpitts oscillator can be written as:

\begin{eqnarray}\label{eq:colpitts}
     \dot{V_1'} &=& \frac{\Phi}{C_1'}[I_L'-\alpha I_s(e^{-\frac{V_2'}{V_T}}-1)], \nonumber \\
    \dot{V_2'} &=& \frac{\Phi}{C_2'}[I_L'+I_s(1-\alpha) (e^{-\frac{V_2'}{V_T}}-1)-\frac{V_{CC}+V_2'}{R_1}], \nonumber \\
     \dot{I_L'} &=& \frac{\Phi}{L'}(V_{CC}-V_1'-V_2'-I_L'R_2).
\end{eqnarray}

\begin{figure}
    \centering
    \includegraphics[width=0.450\textwidth, angle=0]{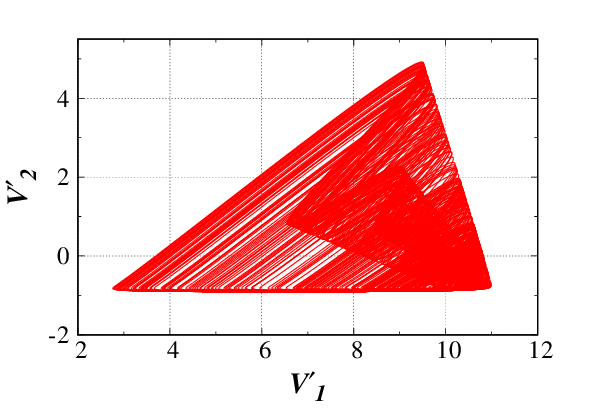}
    \caption{The attractor in the $V'_1-V'_2$ plane for the Colpitts oscillator used to drive the Chua's circuit.}
    \label{fig:Fig4}
\end{figure}

Here $C_1'=C_2'=0.5\ \mu F$, $L'=20\ mH$, $R_1=5000\ \Omega$, $R_2=100\ \Omega$ and $V_{CC}=10\ V$, $I_s=10^{-16}\ A$ is the reverse saturation current of the transistor, $\alpha=0.995$ is the gain of the transistor and $V_T=0.0258\ V$ is the temperature coefficient~\cite{roy2003}. The attractor of Colpitts oscillator in $V'_1-V'_2$ plane is shown in Fig.~\ref{fig:Fig4}. In Eqs.~\eqref{eq:colpitts} we introduce a parameter $\Phi$ to control the time-scale of the driving signal. In circuit realization, this is done by modifying the values of capacitors and inductor as $C_1'\rightarrow \frac{C_1'}{\Phi}$, $C_2'\rightarrow \frac{C_2'}{\Phi}$ and $L'\rightarrow \frac{L'}{\Phi}$.

We fix the circuit parameters to the values mentioned earlier and $R$ = 1920 so that the undriven Chua's circuit is in a chaotic regime with asymptotic dynamics on the double-scroll attractor~\cite{matsumoto1985}. To study the effect of the external chaotic drive on Chua's circuit we fed the voltage $V'_2$ from the Colpitts oscillator through a unidirectional switch, $S$ and a resistance, $R'$ at $V_2$ of the Chua's circuit.

The dynamical equations for the driven Chua's circuit read as

\begin{eqnarray}
    \dot{V_1} &=& \frac{1}{RC_1}{[V_2-V_1-Rf(V_1)]}, \nonumber \\    
    \dot{V_2} &=& \frac{1}{RC_2}(V_1-V_2+R I_L)+ \varepsilon(V_2'-V_2), \nonumber \\
    \dot{I_L} &=& -\frac{V_2}{L}.
\end{eqnarray}

\begin{figure}
    \centering
    \includegraphics[width=0.450\textwidth]{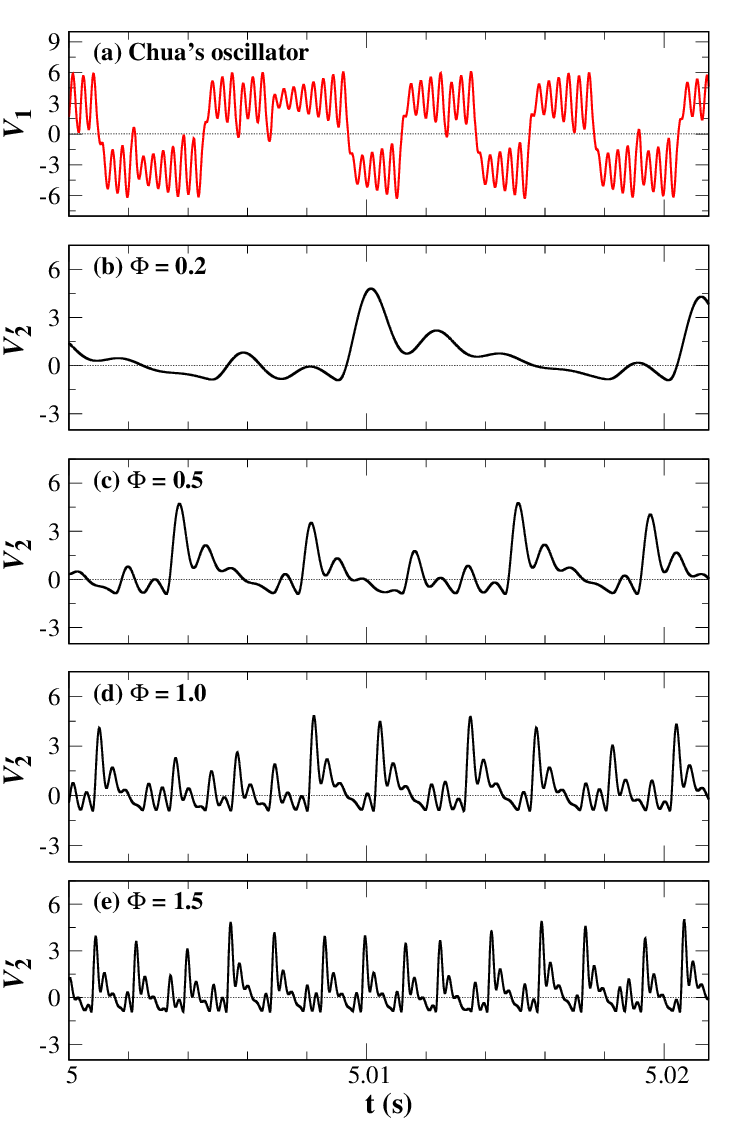}
    \caption{Time-series of (a) $V_1$ of the response Chua's circuit at $R$ = 1920 $\Omega$, and driving signal ($V'_2$) at different values of time-scale parameter, $\Phi$ obtained from circuit simulation.}
    \label{fig:Fig5}
\end{figure}

Here $V'_2$ is used as the driving signal and $\varepsilon$ denotes the driving strength expressed as $\varepsilon=\frac{1}{R'C_2}$. In the circuit, the coupling strength, $\varepsilon$ is varied by changing $R'$ keeping $C_2$ constant. Here we are interested in studying the response of Chua's circuit for different time scales of the drive. The time-series plots for Colpitts oscillators for different values of time-scale parameter, $\Phi$ is plotted in Fig.~\ref{fig:Fig5} to compare it with the time-scale of Chua's oscillator. 

The average time period, $\bar{T}$ of the driving signal is calculated by noting the time duration, $\Delta t$ between consecutive peaks in the time series and taking the average over many such peaks. The inverse of $\bar{T}$ gives the average frequency ($\bar{\nu}$) such that
$$\bar{T}=\frac{\sum_{i=1}^{N} \Delta t_i}{N}$$ and $$\bar{\nu}=\frac{1}{\bar{T}}$$
where $N$ is the total number of peaks in the chosen time interval in the time series.

The average frequency ($\bar{\nu}$) of the $V_1$ variable of the Chua's oscillator in the chaotic regime ($R=1920\Omega$) and the chaotic drive ($V'_2$) for several value of time-scale parameter ($\Phi$) is shown in Table~\ref{tab:tab1} and we note that the average frequency of the driving signal $\bar{\nu}$ varies linearly with the time-scale parameter $\Phi$. 
\begin{table}[H]
    \centering
    \begin{tabular}{|c|c|c|}
    \toprule
    \hline 
        $\Phi$ & Average time period, & Average frequency,\\
         &$\bar{T}$ (s) & $\bar{\nu}$ (Hz)\\
        \hline
         0.2& $2.427\times 10^{-3}$ & 412\\
         0.5& $9.8\times 10^{-4}$ & 1021\\
         1.0& $4.923\times 10^{-4}$ & 2031\\
         1.5& $3.238\times 10^{-4}$ & 3088\\
         \hline
         Chua's Circuit& $3.370\times 10^{-4}$ &2967 \\
	\hline
	\bottomrule   
    \end{tabular}
    \caption{Comparison of average frequency of Chua's oscillator and the frequency of chaotic drive for different values of time-scale parameter $\Phi$.}
    \label{tab:tab1}
\end{table}

\begin{figure}
    \centering
    \includegraphics[width=0.50\textwidth]{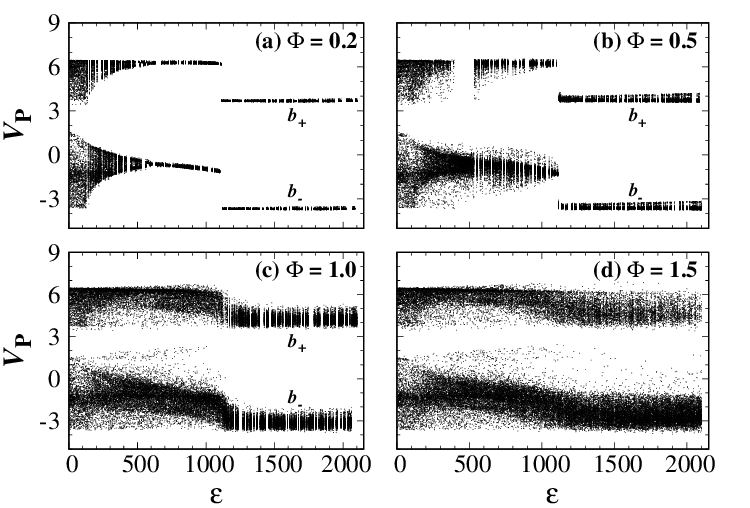}
    \caption{ Bifurcation diagram of the driven Chua's circuit with increasing coupling strength, $\varepsilon$ for different values of time scale parameter, $\Phi$ taking random initial conditions for each $\varepsilon$.}  
    \label{fig:Fig6}  
\end{figure}

\begin{figure}
    \centering
    \includegraphics[width=0.50\textwidth]{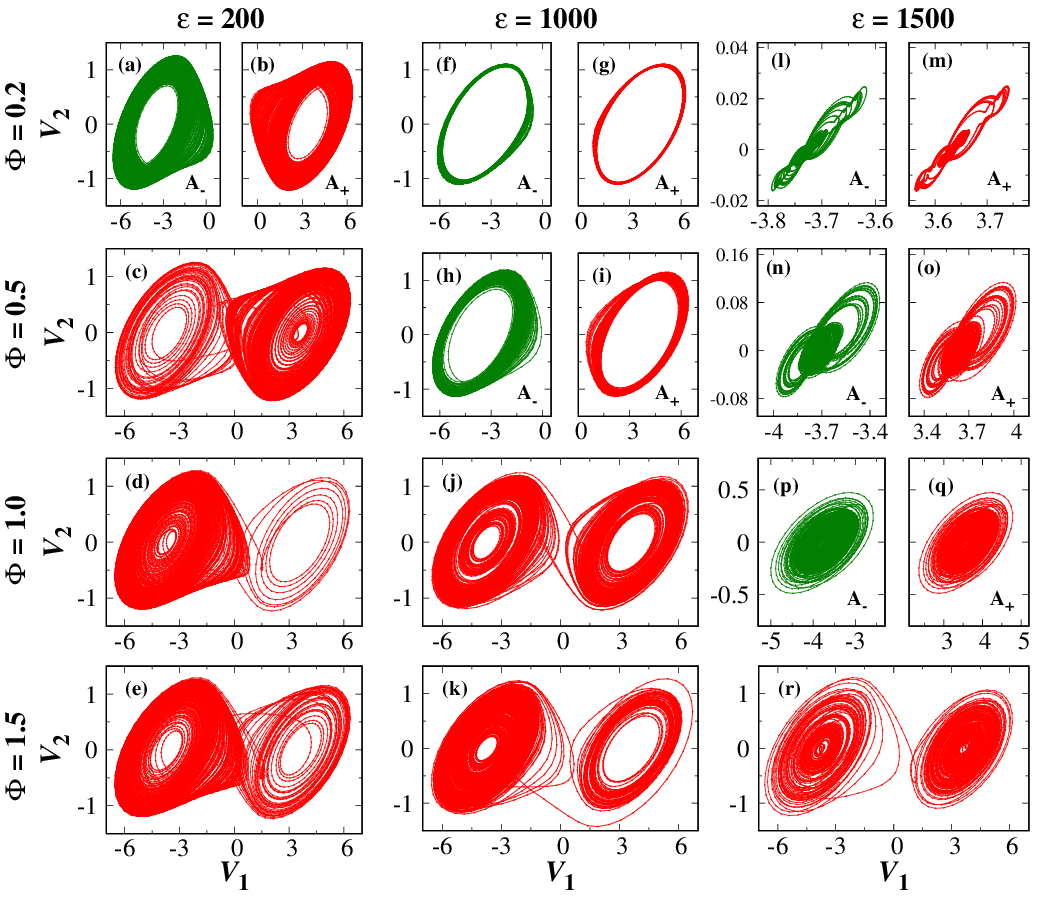}
    \caption{Attractors of driven Chua's circuit are plotted for different values of time-scale parameter, $\Phi$ and driving strength, $\varepsilon$.}   
    \label{fig:Fig7} 
\end{figure}

The bifurcation diagram for the driven Chua's circuit with increasing $\varepsilon$ gives an overview of the possible attractors. In Fig.~\ref{fig:Fig6} we have plotted the peak voltage of $V_1$ denoted by $V_p$ for a random initial condition for each $\varepsilon$. Here we observe that when the driving signal is much slower than the response indicated by small value of $\Phi$ (Fig.~\ref{fig:Fig6}(a)), the asymptotic dynamics of the Chua's circuit is on a double-scroll chaotic attractor when the driving strength, $\varepsilon$ is low. As $\varepsilon$ increases (for fixed $\Phi$) the attractor changes from double-scroll to single-scroll. As we can see in Fig.~\ref{fig:Fig6}(a) there are two variants of single-scroll chaotic attractors corresponding to the upper ($b_+$) and lower ($b_-$) branches in the bifurcation diagram, for $b_+$ the attractor has positive $V_1$ values and is denoted by $A_+$ whereas for $b_-$ the attractor has negative $V_1$ values and refereed as $A_-$. We also observe that the size of these single-scroll chaotic attractors ($A_+$ and $A_-$) decreases with increase in driving strength $\varepsilon$ as evident from the attractors plotted in Fig.~\ref{fig:Fig7}. On the contrary when the time-scale of drive is comparable to that of response, the long time dynamics of the response always remains on the double-scroll chaotic attractor only (Fig.~\ref{fig:Fig6}(d)). The shape of the attractors of the driven Chua's oscillator for different values of $\Phi$ and $\varepsilon$ are shown in Fig.~\ref{fig:Fig7} showing bistability in the parameter region where the single-scroll attractors exist.

\begin{figure}
    \centering
    \includegraphics [angle=270, width=0.40\textwidth]{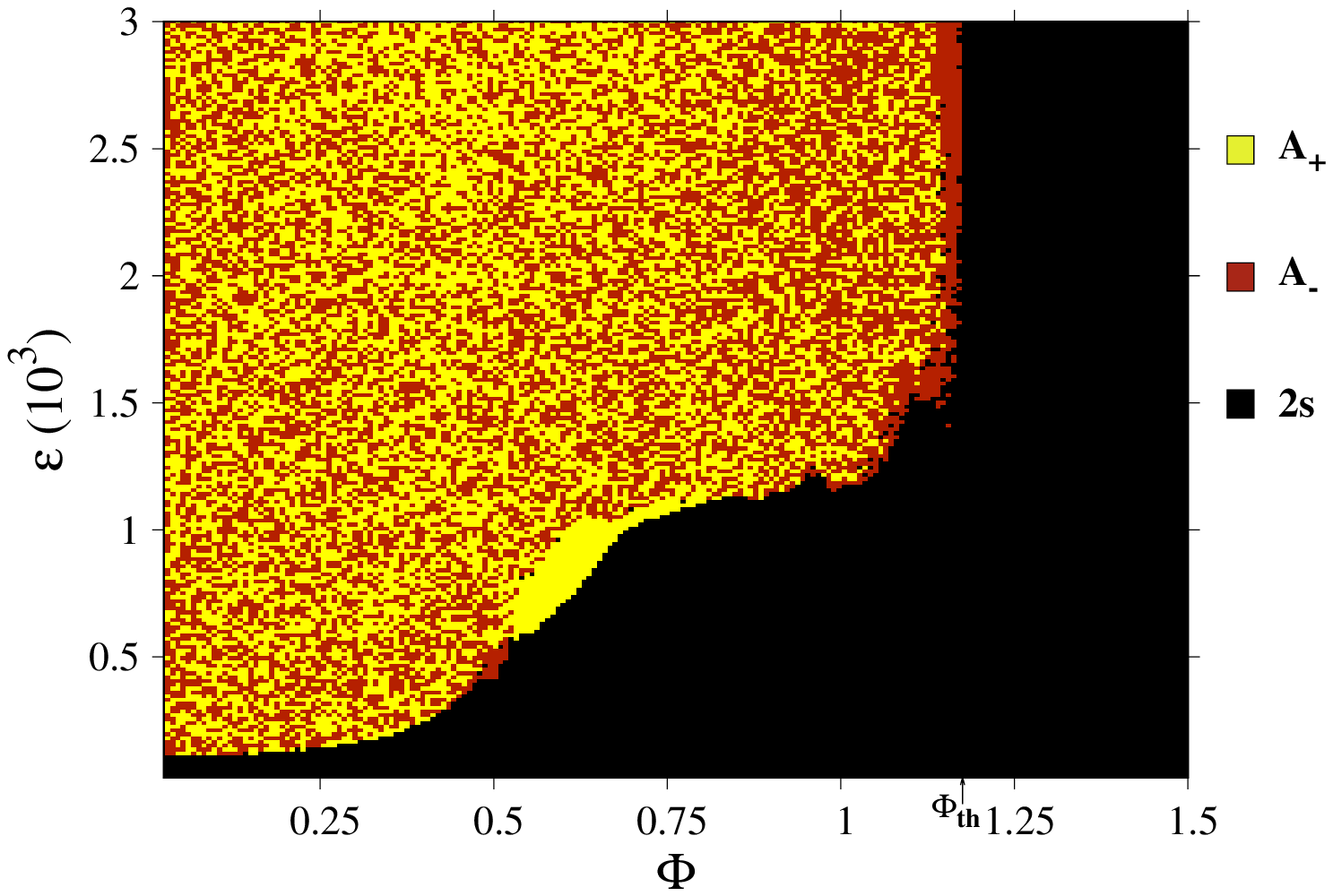}
    \caption{Figure showing existence of double-scroll (2s) and single-scroll ($A_+$ and $A_-$) attractors for the driven Chua's circuit in $\Phi$-$\varepsilon$ parameter space with initial conditions taken randomly~\cite{sim-details}.}
    \label{fig:Fig8}   
\end{figure}

We also study the transition of the attractor from double-scroll to single-scroll in the parameter space of $\Phi$ and $\varepsilon$. To distinguish between single-scroll and double-scroll attractors, we consider the time-series of $V_1$ after removing the transients, take maxima, $V_{1(max)}$ and minima, $V_{1(min)}$ of the time series. We compute the quantity $\chi$ defined as $\chi = V_{1(max)} + V_{1(min)}$. If $\chi \approx 0$ the dynamics of the Chua's oscillator is on the double-scroll attractor whereas $\chi$ positive ($>$ 5) and negative ($<$-5) indicate that the trajectories converge to $A_+$ and $A_-$ attractors respectively. As seen in Fig.~\ref{fig:Fig8} the transition from double-scroll to single-scroll attractor occurs at a lower value of $\varepsilon$ when the drive is much slower than the response (low value of $\Phi$). As $\Phi$ increases the point of transition moves towards higher $\varepsilon$. When the time scale of the drive approaches that of the response, there is no transition from double-scroll to single-scroll chaotic attractor and we observe double-scroll attractors in the entire $\varepsilon$ range. This threshold value of $\Phi$ beyond which only double-scroll attractors are observed is marked as $\Phi_{th}$  ($\approx 1.18$) in Fig.~\ref{fig:Fig8}. From these observations we conclude that a slow driving signal favours the existence of single-scroll chaotic attractors and hence bistability in the driven Chua's circuit.

\section{Generalized synchronization in driven Chua's circuit} \label{sec:gs}

Non-identical unidirectionally coupled systems can achieve synchronization in a generalized manner wherein the dynamical variables of the response system can be expressed as a function of the dynamical variables of the drive~\cite{rulkov1995}. One way to study the generalized synchrony between response and drive is to employ the auxiliary system approach \cite{abarbanel1996} where an identical copy of the response system, known as the auxiliary system, is considered and the common driving signal is fed to both response and auxiliary units. If the response and auxiliary systems are completely synchronized then the response is said to be in generalized synchrony with the drive. 

In order to study the generalized synchrony between the Chua's circuit and the chaotic drive, we consider an identical Chua's system as an auxiliary with dynamical variables denoted by $V^a_1$, $V^a_2$, and $I^a_L$ where the superscript `$a$' is used to distinguish the state variables of the auxiliary system from that of the response Chua's circuit.
 
The asymptotic dynamics of the two Chua's circuits (response and auxiliary) are on the double-scroll chaotic attractor (fixing $R=1920\  \Omega$) when no driving signal is applied. To investigate the nature of generalized synchrony between drive and Chua's oscillator, we drive response and auxiliary systems by a common chaotic signal $V_2'$ of the common base Colpitts oscillator.

\begin{figure}
    \centering
    \includegraphics[width=0.50\textwidth]{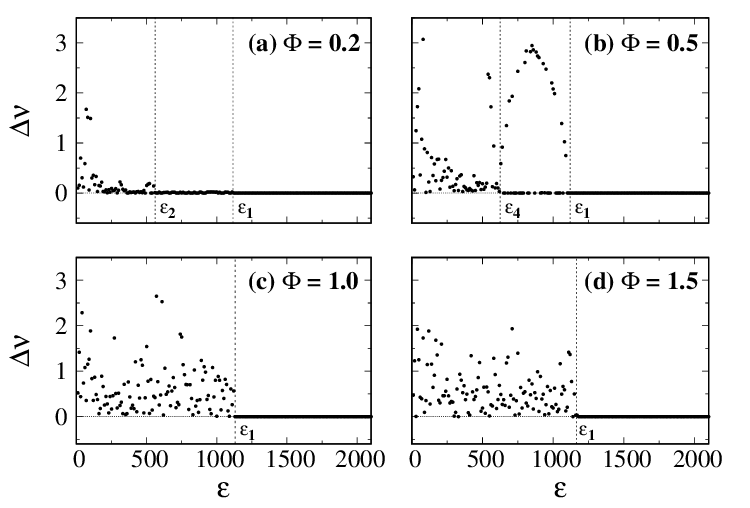} 
    \caption{Plot of difference in average frequency between the response and auxiliary Chua's oscillator, $\Delta \nu$ as a function of driving strength, $\varepsilon$ for different values of $\Phi$ taking random initial conditions at each $\varepsilon$.}   
    \label{fig:Fig9}  
\end{figure}

The synchronization between the response and auxiliary units is studied as a function of coupling strength, $\varepsilon$ for a fixed value of time-scale parameter $\Phi$. 
As $\varepsilon$ increases (for fixed $\Phi$) the trajectories of the two Chua's oscillators transition from being desynchronized to a synchronized state at a coupling threshold marked as $\varepsilon_1$. To identify this threshold we plot the difference of average frequency, $\Delta \nu$ = $|{\bar{\nu}-\bar{\nu}^a}|$ of the two Chua's oscillators as a function of $\varepsilon$ for different $\Phi$ in Fig.~\ref{fig:Fig9}. When the time-scale of drive is much slower than the response the transition from desynchronized ($\Delta \nu \neq 0$) to the complete synchronized state ($\Delta \nu = 0$) is through a region of multistability where synchronized state coexist with other state for which $\Delta \nu \approx 0$ ((between $\varepsilon_1$ and $\varepsilon_2$) shown in Fig.~\ref{fig:Fig9}(a). As $\Phi$ increases the difference in average frequencies increases and there is a region in $\varepsilon$ space where synchronized state exist with the desynchronized one indicated by scattered points for intermediate values of $\varepsilon$ (region between $\varepsilon_1$ and $\varepsilon_4$) in Fig.~\ref{fig:Fig9}(b). This multistability of synchronization and desynchronization between response and auxiliary units disappears as the time-scale of drive increases and becomes equivalent to the time-scale of the response (Fig.~\ref{fig:Fig9}(c)-(d)).

\begin{figure}
    \centering
    \includegraphics[width=0.50\textwidth]{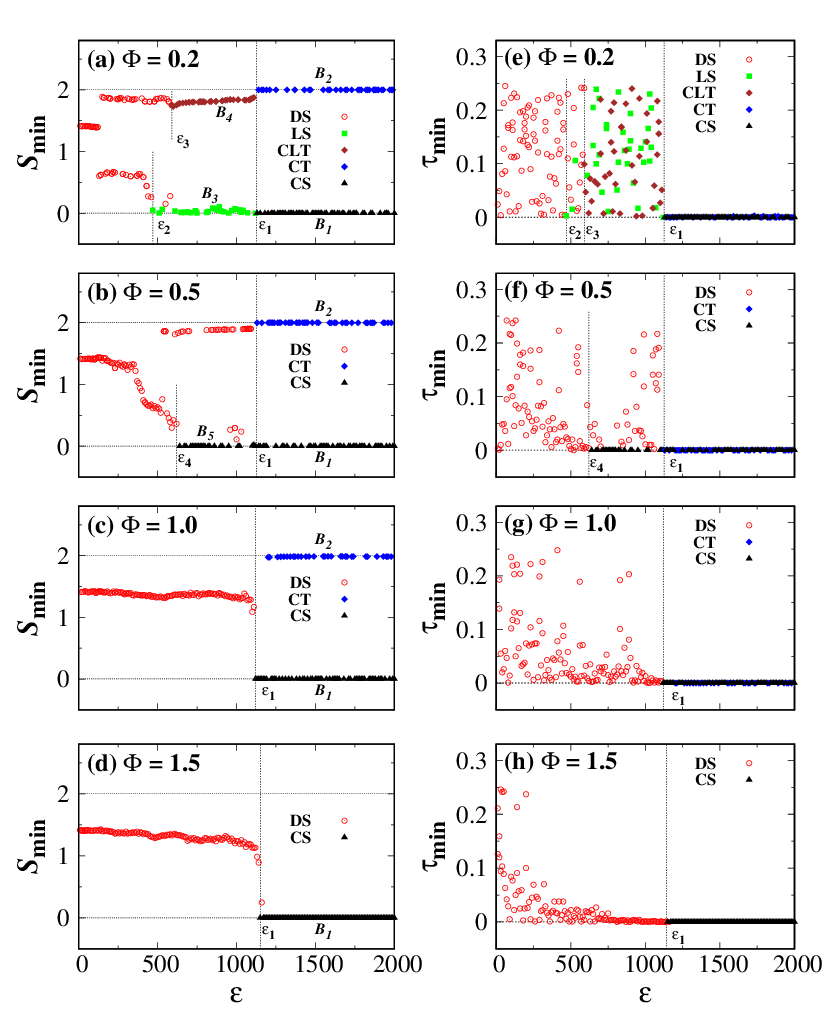}
    \caption{The minimum of similarity parameter $S_{min}$ (left panel) and the corresponding $\tau_{min}$ (right panel) with increasing $\varepsilon$ for different values of time-scale parameter, $\Phi$ showing complete synchronization (CS with black triangles),  correlated trajectories (CT with blue diamonds), lag synchronized state (LS with green squares), correlated lag trajectories (CLT with brown filled diamonds), and desynchronized state (DS with red circles). }      
    \label{fig:Fig10}  
\end{figure}

In order to probe further into the features of the emerging states as $\Phi$ and $\varepsilon$ vary, we calculate the similarity parameter~\cite{rosenblum1997} between the time series of the state variables, $V_1$ and $V^a_1$ of the response and the auxiliary respectively. The similarity parameter  $S(\tau)$ is defined as 

\begin{equation}
S(\tau)=\sqrt{{\frac{\langle[ V^a_{1}(t+\tau)-V_{1}(t)]^2\rangle}{[\langle (V_{1}(t))^2\rangle\langle (V^a_{1}(t))^2\rangle]^\frac{1}{2}}}},
\end{equation}
where $\tau$ is the time shift considered between the time-series of the two driven Chua's circuits and $\langle . \rangle$ denote the time averages. The minimum value of similarity parameter, $S_{min}$ and the corresponding $\tau_{min}$ are plotted in Fig.~\ref{fig:Fig10} as a function of driving strength $\varepsilon$ for different values of time-scale parameter $\Phi$. We observe that depending upon the values of $\Phi$, $\varepsilon$ and initial conditions for the state variables, the two Chua's oscillators can be in one of the following states: 

\begin{enumerate}[label=(\roman*)]

\item Complete synchronization (CS) where
 
\begin{eqnarray}\label{eq:cs}
	V_1(t) &=& V^a_1(t), \nonumber  \\
   V_2(t) &=& V^a_2(t), \nonumber \\
   I_L(t) &=& I^a_L(t) .
\end{eqnarray}

\item Lag synchronization (LS) where

 \begin{eqnarray}\label{eq:ls}
 V_1 (t) &=& V^a_1(t + \Delta t), \nonumber  \\
  V_2(t) &=& V^a_2(t + \Delta t), \nonumber  \\
 I_L (t) &=& I^a_L(t + \Delta t).
  \end{eqnarray}
Here $\Delta t$ is the time lag between the time-series of the state variables of response and auxiliary systems.

\item Correlated trajectories (CT) where

 \begin{eqnarray}\label{eq:ct}
 V_1(t) &=& V^a_1(t)+k_1, \nonumber \\
  V_2(t) &=& V^a_2(t), \nonumber \\
 I_L(t) &=& I^a_L(t)+k_2.
 \end{eqnarray}
 $k_1$ and $k_2$ are time independent constants. These constants are the same for all $\epsilon$ for a fixed $\Phi$.
  
  \item Correlated lag trajectories (CLT) where
    
 \begin{eqnarray}\label{eq:clt}
 V_1(t) &=& -V^a_1(t + \Delta t), \nonumber \\
  V_2(t) &=& -V^a_2(t + \Delta t), \nonumber \\
 I_L(t) &=& -I^a_L(t +\Delta t). 
 \end{eqnarray}
  
  Here $\Delta t$ is the time lag. In CLT state the response and the auxiliary systems are in anti-phase synchrony and follow Eq.~\eqref{eq:clt} for most times.
  
 \item Desynchronized state (DS) where there is no correlation between the state variable of the response and auxiliary Chua's oscillators. 

\end{enumerate}
 
For CS $S_{min}=0$ at $\tau=0$ whereas $S_{min}\simeq 0$ for $\tau \neq 0$ represents LS. For DS, the similarity parameter $S_{min}>>0$ for all $\tau$ values ~\cite{rosenblum1997}. In order to differentiate and verify the existence of different states we modify the similarity parameter as:

\begin{equation}
S_{mod}(\tau) = \sqrt{{\frac{\langle[ (V^a_{1}(t+\tau) + k) \pm V_{1}(t)]^2\rangle}{[\langle (V_{1}(t))^2\rangle\langle (V^a_{1}(t))^2\rangle]^\frac{1}{2}}}}.
\end{equation}

For the CS, LS, and CLT states $k$=0 while $k$ is finite for CT state. For CLT state we consider the (+) sign to verify the relation between the state variables (Eq.~\eqref{eq:clt}) whereas (-) sign applies for other states.

\begin{figure*}
    \centering
    \includegraphics[width=0.7\textwidth]{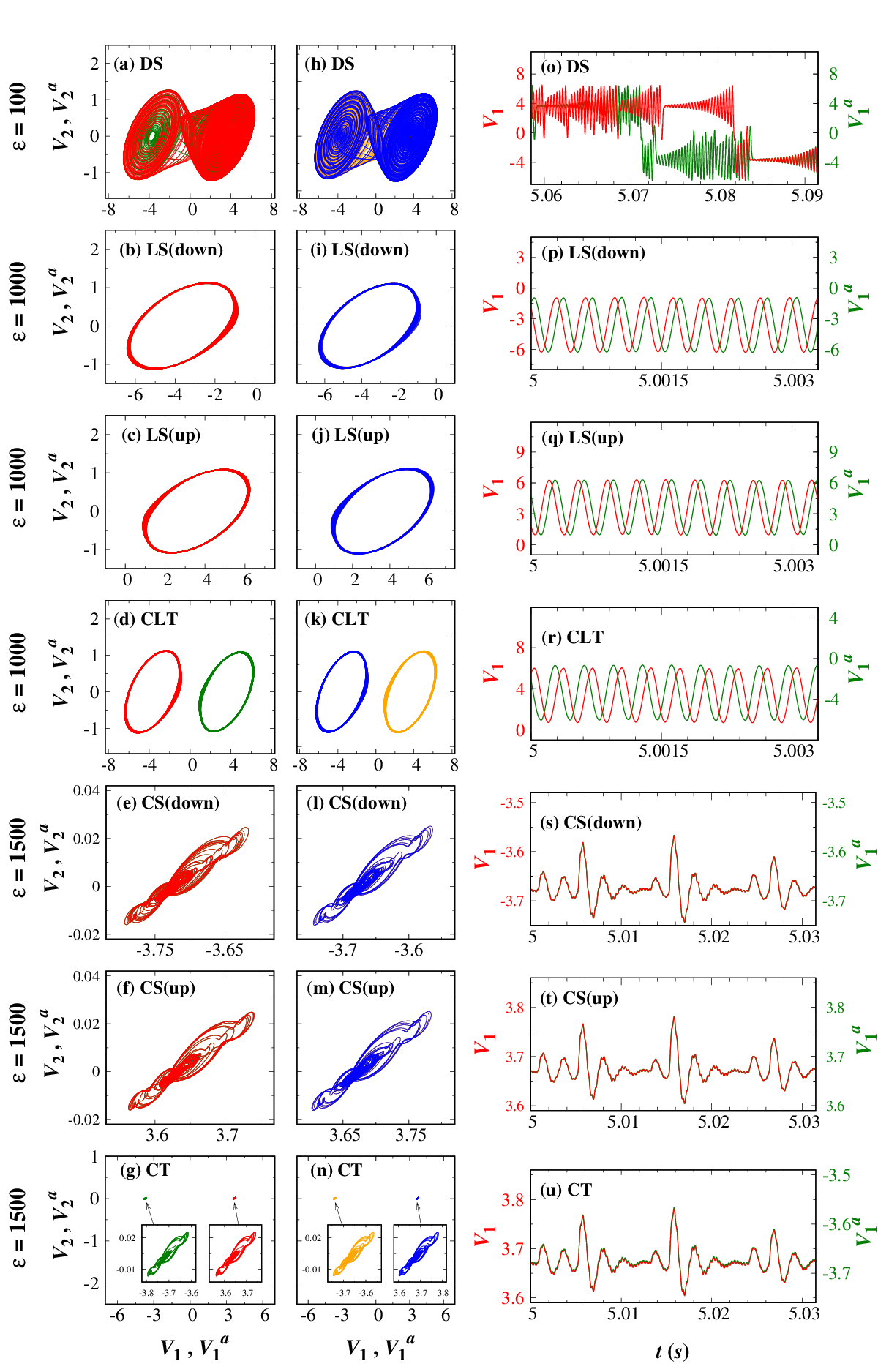}
    \caption{Attractors of the response (in red) and auxiliary (in blue) Chua's circuits obtained from circuit simulation in Multisim [right panel (a)-(g)] and numerically [middle panel(h)-(n)] for various observed states DS, LS, CLT, CS and CT for $\Phi$ = 0.2. The right panel shows the time-series plots of $V_1$ of response (red color) and $V^a_1$ (blue color) of auxiliary obtained numerically.}
    \label{fig:Fig11}    
\end{figure*}

From Fig.~\ref{fig:Fig10} we observe that when driving is much slower than response ($\Phi=0.2$), for low values of driving strength, $\varepsilon$ the two Chua's oscillators are in DS indicating no GS between the Chua's circuit and the drive. As $\varepsilon$ increases and crosses a threshold value we see two branches $B_3$ ($\varepsilon_2 < \varepsilon < \varepsilon_1$ ) and $B_4$ (after $\varepsilon_3 < \varepsilon < \varepsilon_1$) emerging from the desynchronized region, $B_4$ corresponds to the CLT state where response and auxiliary go to two different attractors ($A_+ A_-$ or $A_-A_+$) and also have a time lag while $B_3$ represents LS between the two Chua's oscillators.
These two branches indicate different scenarios of \emph {remote} synchronization (RS) where two Chua's circuits not directly coupled to each other synchronize (indicated by equal average frequency) via a common drive but have different phase relation between their dynamical variables. Therefore one can anticipate some relationship between the dynamics of response Chua's oscillator and the drive. We call the RS between the two Chua's oscillators corresponding to the two branches $B_3$ and $B_4$ as RS of type III (RS-III) and RS of type IV (RS-IV) respectively. As $\varepsilon$ increases further, exceeding $\varepsilon_1$ the lag synchronized branch ($B_3$) evolves to complete synchrony (CS) denoted by $B_1$ while the CLT changes to CT state ($B_2$ branch) where the two Chua's oscillators go to two different single-scroll attractors ($A_+$ or $A_-$) with no time lag but there is a correlation between the state variables of the two oscillators given in Eq.~\eqref{eq:ct}. The state CT seems to indicate a new kind of RS between the drive and the response systems, which we name as remote synchronization of type II (RS-II). Therefore in addition to the case when the response and auxiliary systems are in complete synchrony (CS), say RS-I, which indicates generalized synchrony between the drive and response, we observe new scenarios of \emph {remote} synchronization namely RS-II, RS-III and RS-IV where the trajectories of response and auxiliary Chua's oscillators are correlated in different manner. Moreover, in Fig.~\ref{fig:Fig10}(a) we observe two regions of multistability: one at intermediate values of $\varepsilon$ (between $\varepsilon_3$ and $\varepsilon_1$) where RS-III coexist with RS-IV (LS coexists with CLT) and other at higher $\varepsilon$ where RS-II coexist with RS-I (CS coexists with CT). The later region appears more interesting as it shows the coexistence of remote synchronization with generalized synchronized states.

For higher values of time-scale parameter, $\Phi$ the lag between the time-series of the Chua's oscillators decreases and LS state converts into CS while the CLT branch becomes desynchronized (Fig.~\ref{fig:Fig10}(b)). This gives rise to a multistable region between $\varepsilon_4$ and $\varepsilon_1$ where CS and DS coexist and one can expect chimera-like states~\cite{kuramoto2002, abrams2004} to emerge in an ensemble of Chua's oscillators driven by a common chaotic signal. Therefore as $\Phi$ increases the LS and CLT states disappear, and the bistability is between CS and CT states only for $\varepsilon > \varepsilon_1$ (Fig.~\ref{fig:Fig10}(c)). Further increasing $\Phi$ makes the CT state vanish and when the time-scale of the drive is equivalent to that of the response ($\Phi \approx 1.46$), only CS exists above a critical coupling, $\varepsilon_{1}$. This investigation indicates that a significantly slower drive induces novel types of RS and multistability in the drive-response system which disappears when time-scale of drive is comparable to that of the response.        

The attractors (left panel Fig.~\ref{fig:Fig11}(a-g)) and the time-series (right panel Fig.~\ref{fig:Fig11}(o-u)) plots obtained from circuit simulation for these different states are shown. To illustrate the agreement between results obtained from circuit simulations and numerical simulation of model equations the attractors obtained numerically are shown in the middle panel Fig.~\ref{fig:Fig11}(h-n). We observe that for low $\Phi$ the synchronization of response and auxiliary is on single-scroll attractor whereas for high $\Phi$ the synchronization is on double-scroll attractor.  
For the lower frequency of the drive, CS can be on two types of single-scroll attractors: one with positive $V_1$ and $V^a_1$ values denoted by CS(up) and the other with negative $V_1$ and $V^a_1$ values marked as CS(down) shown in Fig.~\ref{fig:Fig11}. Also the size of the single-scroll attractor on which dynamics is synchronized decreases with increase in $\varepsilon$. For higher $\Phi$, LS and CLT are first to disappear and then CT disappears when complete synchronization of the two Chua's oscillators is on the double-scroll attractor indicated by CS(2s).

\begin{figure}
    \centering
    \includegraphics[width=0.50\textwidth]{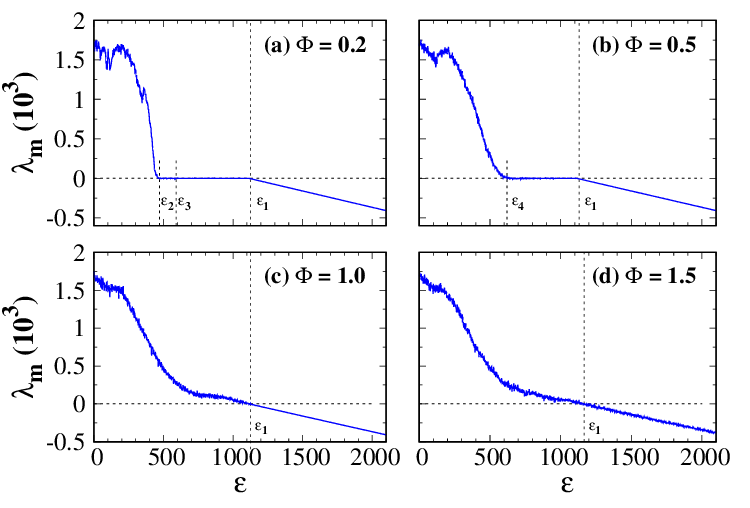}
    \caption{Maximum conditional Lyapunov exponent, $\lambda_m$ with varying $\varepsilon$ for different values of time-scale of the drive, $\Phi$.}  
    \label{fig:Fig12}  
\end{figure}

\begin{figure}
    \centering
    \includegraphics [angle=0, width=0.48\textwidth]{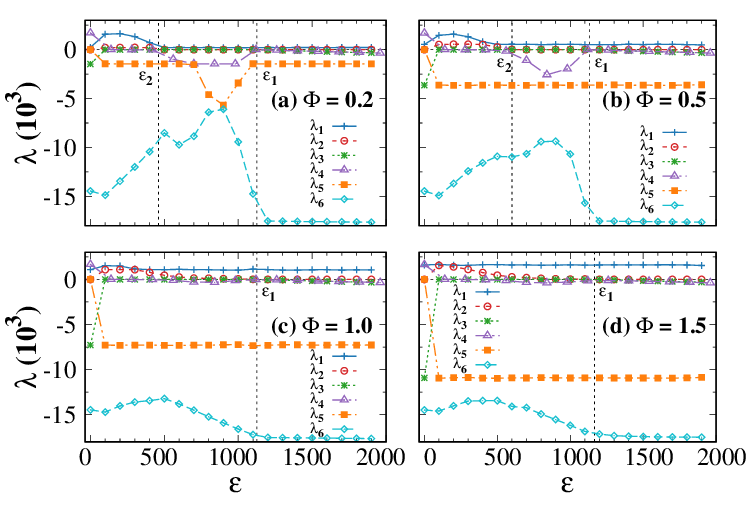}
    \caption{All six Lyapunov exponents of the driven system are plotted with the variation of $\varepsilon$ for different values of $\Phi$.}
    \label{fig:Fig13}   
\end{figure}

In order to analyse the stability of the synchronized states we compute the conditional Lyapunov exponent for the driven Chua's circuit. In Fig.~\ref{fig:Fig12}, the largest conditional Lyapunov exponent, $\lambda_m$ is plotted as a function of coupling strength, $\varepsilon$ for different values of $\Phi$. We observe that for low values of $\varepsilon$, $\lambda_m$ is positive suggesting incoherence between the two Chua's oscillators and hence no GS between drive and response. After driving strength crosses a threshold $\varepsilon_1$, $\lambda_m$ becomes negative indicating GS between drive and the response systems~\cite{pecora1990}. The value of $\lambda_m \approx 0$ when there is lag synchronization (LS). In the region where RS-III coexist with RS-IV, $\lambda_m$ has either small positive or negative values depending upon the initial conditions. From Fig.~\ref{fig:Fig12}(c)-(d) we notice that for higher values of $\Phi$ the desynchronized region expands and we are unlikely to get LS and CLT states. We also note that $\lambda_m$ is negative in the parameter region where RS-I and RS-II coexist. The Lyapunov exponent spectrum for the driven system is also plotted (see  Fig.~\ref{fig:Fig13}). The drive-response system will have six Lyapunov exponents (LEs). We observe that for low coupling $\varepsilon$ the drive-response system is hyperchaotic characterized by two positive LEs. One LE is always positive due to the chaotic nature of the drive. At $\varepsilon_1$ and $\varepsilon_2$ one or more LEs come close to zero (avoided crossing) indicating change in system's dynamics (response in this case since drive remains unaffected by the coupling). If we separate the LEs corresponding to the drive we will be left with three LEs that represent the dynamical nature of the response. Doing this we found that the largest LE of the response system is close to zero in the range $\varepsilon_2 <\varepsilon <\varepsilon_1$ whereas it is negative when $\varepsilon>\varepsilon1$.

\begin{figure}
    \centering
    \includegraphics[width=0.50\textwidth]{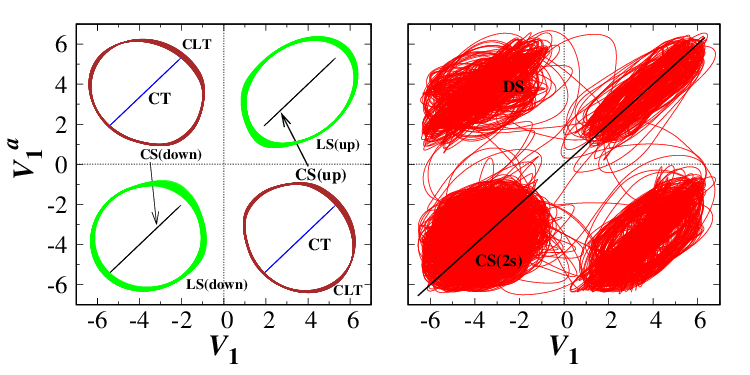}
    \caption{$V_1$ of response vs $V^a_1$ of auxiliary for the observed states, on the left CS(up), CS(down), LS(up), LS(down) CT, CLT and on the right DS (red color) and CS(2s) (Black color). For LS and DS $\Phi$ = 0.2 and $\epsilon$ = 1000; for CT and CS(up/down) $\Phi$ = 1.0 and for CS(2s) $\Phi$ = 1.5 at $\epsilon$ = 1500.  }        
    \label{fig:Fig14}
\end{figure}

To confirm the generalized synchronization (GS) between the response Chua's circuit and the drive, we plot $V_1^a$ vs $V_1$ in Fig.~\ref{fig:Fig14}. $V_1^a$ varies linearly with $V_1$ for CS and CT indicating GS between drive and response. When the synchronization is on the single-scroll attractor the plots for CS lie either in first or third quadrant while plots for CT lie in second or fourth quadrants in the $V_1$-$V_1^a$ space. At higher $\Phi$, when the synchronization is on double-scroll attractor only, the plots are in the form of straight line stretching through first and third quadrants and passing through origin. For DS the variation of $V_1^a$ with respect to $V_1$ is random whereas the plots look like a closed orbits for CLT and LS states indicating some correlation between the trajectories of response and auxiliary units.

\begin{figure}
    \centering
    \includegraphics[width=0.50\textwidth]{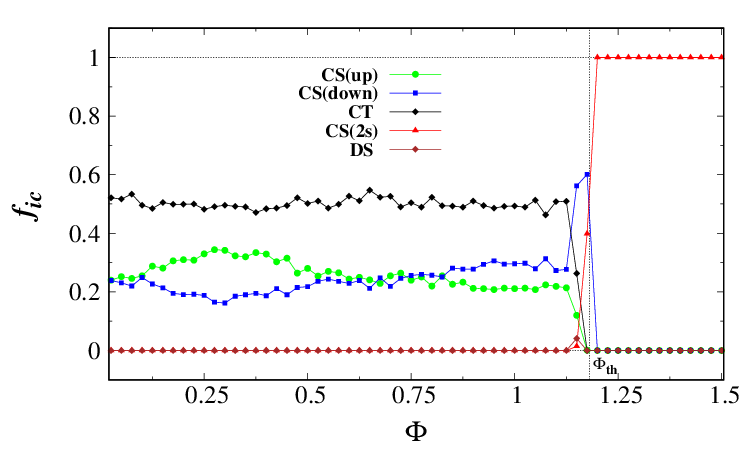}
    \label{fig:enter-label}
    \caption{Fraction of initial conditions, $f_{ic}$ leading to different synchronized states as a function of time-scale parameter $\Phi$ of the driving signal for fixed coupling ($\varepsilon = 1500$) considering $10^3$ random initial conditions at each $\Phi$.}    
    \label{fig:Fig15}   
\end{figure}

Since the driven Chua's circuit exhibits multistability, we examine the fraction of initial conditions leading to various attractors.  
For a fixed $\varepsilon$, we take $10^3$ initial conditions randomly and compute the fraction of initial conditions, $f_{ic}$ going to the various states: CS(up), CS(down), CS(2s) and CT (see Fig.~\ref{fig:Fig15}). The figure confirms that when $\Phi$ is below a certain value, $\Phi_{th}$ the synchronization between response and auxiliary is on the single-scroll attractors marked as CS(up) (when the synchronization is on $A_+$) and CS(down) (when the synchronization is on $A_-$) or there could be scenario where one oscillator asymptote to $A_+$ and the other goes to $A_-$ or vice-versa termed as CT state. When $\Phi > \Phi_{th}$, the two Chua's oscillators synchronize on to the double-scroll attractors (CS(2s)). Therefore, it can be concluded that when the drive has lower frequency as compared to the response, we expect synchronization on single-scroll attractor and hence multistability of various kinds while when the frequency of drive is comparable to the response the synchronized dynamics is on double-scroll attractor only and hence no multistability.

\section{Analysis of dynamics in the $\lambda_{m}$ = 0 region } \label{sec:inter}

In order to understand the dynamical nature of Chua's circuit in the region where $\lambda_m \approx 0$, we analyze the time series of the response system at different $\Phi$ and fixed coupling $\epsilon$ using various measures such as 0-1 chaos test, Fourier transform (FFT) and recurrence plots. 

\subsection{0-1 Chaos test}
0-1 test is designed to distinguish chaotic behaviour from regular dynamics in deterministic systems. If the outcome of the test is 1 it indicates chaotic nature of the system whereas 0 result corresponds to regular dynamics. For the implementation of 0-1 test we follow the steps outlined in the Ref. \cite{gottwald2009}. We consider the time series of $V_1$ of length $N$ and define translation variables $p(n)$ and $q(n)$ as follow: 
 
 \begin{eqnarray}\label{eq:chaostest_pq}
 	p(n) &=& \sum_{j=1}^{n}\eta(j)\cos(jc) , \nonumber \\   
 		q(n) &=& \sum_{j=1}^{n}\eta(j)\sin(jc), 
 \end{eqnarray}
 
where ${\eta} = 1, 2, \dots{N}$ and $\eta(j)$ is a observable constructed from the time series.\\
The value of $c$ can be chosen from the interval (0,$\pi$). The behavior of these translational variables determine whether the dynamics is chaotic or regular. 
To determine the diffusive or non-diffusive behaviour of $p$ and $q$ we calculate the mean square displacement ${M(n)}$ as

  \begin{equation}\label{eq:chaostest_m}
  	M(n) = \lim_{N\to\infty}\frac{1}{N}\sum_{j=1}^{N}[p(j+n)-p(j)]^2 +[q(j+n)-q(j)]^2.    
  	\end{equation}
  
For  large $N$, we require $n \ll N$. Hence we calculate $M(n)$ for $n \leq N_c$ such that $N_c \ll N$. We take $N_c$ = $N/10$ for our analysis. For better convergence of $M(n)$, a modified mean square displacement $D(n)$ can be used defined as,

   \begin{equation}
   	D(n) = M(n) - V_{osc}(c,n)
   \end{equation}

where

    \begin{equation}\label{eq:chaostest_v}
   	V_{osc}(c,n) = \lim_{N\to\infty}\frac{1}{N}\sum_{j=1}^{N}\eta(j)^2\frac{1-\cos(nc)}{1-\cos(c)}. 
   \end{equation}

We compute the asymptotic growth rate $K$ for a particular value of $c$ ($c$ = 2) from modified mean square displacement $D(n)$ by correlation method in which $K$ is defined as the correlation coefficient of the vectors
 
 \begin{equation}\label{eq:K}
  \zeta =  (1,2,....., N_c) \nonumber 
  \end{equation}
and
  \begin{equation}\label{eq:Kc_1}
 	\Delta = (D(1), D(2), D(3),.....,D(N_c))  
  \end{equation}
such that
\begin{equation}\label{eq:K_2}
	K = corr(\zeta, \Delta) = \frac{cov(\zeta, \Delta)}{\sqrt{var(\zeta)var(\Delta)}} \in [-1,1] 
\end{equation}
 
where variance and covariance are defined in usual manner:
   
   \begin{equation}\label{eq:K_3}
   	\text{cov}(x_1,x_2) =\frac{1}{q}\sum_{j=1}^{q}(x_1(j)- \bar{x_1})(x_2(j)-\bar{x_2}) \nonumber
	   \end{equation}

	 with, $\bar{x_1} = \frac{1}{q}\sum_{j=1}^{q} x_1(j)$
  
   and
    \begin{equation}\label{eq:K_4}
   	\text{var}(x_1) = \text{cov}(x_1,x_1). 
   \end{equation}

Further 100 random choices of $c$ in the interval (0,$\pi$) are taken and $K$ is calculated for each $c$. The median of these $K$ values are obtained as $K_m$. The obtained values of $K$ at $c$ = 2 and $K_m$ for time series of $V_1$ at different $\Phi$ is tabulated in Table ~\ref{tab:tab2}. The value of $K_m$ close to zero indicates that the dynamics is regular and close to one reflects chaotic dynamics. Therefore,  the values of $K_m$ suggest that as $\Phi$ increases from 0.2 to 1.5 the dynamics of Chua's circuit changes from regular to chaotic.

\begin{table}[h]
	\centering
	\setlength{\tabcolsep}{12pt} 
	\begin{tabular}{|c|c|c|}
		\hline
		\bm{$\Phi$} & \bm{$K$}& \bm{$K_m$}\\ \hline
		0.2 & -0.01149 & 0.01121 \\ \hline
		0.5 & 0.03480 & 0.18056 \\ \hline
		1.0 & 0.85941  & 0.97902  \\ \hline
		1.5 & 0.98545  & 0.99274 \\ \hline
	\end{tabular}
	\caption{Asymptotic growth rate, $K$ at $c$ = 2 and the median of $K$, $K_m$ are shown for different values of $\Phi$. The coupling parameter is fixed to $\varepsilon$=1000.}
	\label{tab:tab2}
\end{table}

\begin{figure}
	\centering
	\includegraphics[width=0.50\textwidth]{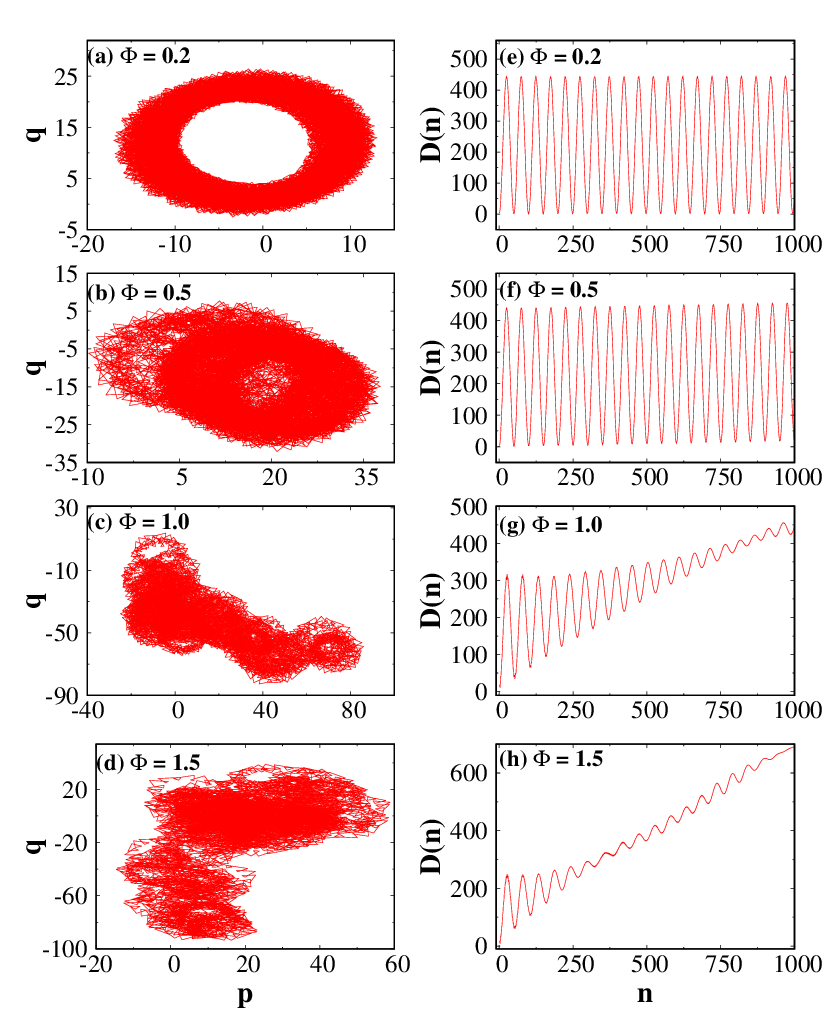}
	\label{fig:enter-label}
	\caption{ The $p-q$ phase portraits are shown in the left panel [(a)-(d)] with their respective modified mean square displacement, $D(n)$ in the right panel [(e)-(h)] for different values of $\Phi$ at fixed coupling, $\varepsilon$ = 1000. The system is simulated for $10^6$ time steps with the sampling  interval $T$ = 100 and $N$=10$^4$. The plots are shown for $c$ = 2.  }    
	\label{fig:Fig16}   
\end{figure}

We also obtain $p$ vs $q$ plots for different $\Phi$ shown in Fig.~\ref{fig:Fig16}. When $p-q$ phase space is bounded the dynamics of the underlying trajectory is regular \cite{gopal2013} which either can be periodic or quasiperiodic. Also, the modified mean square displacement $D(n)$ is confined within a range for $\Phi$ = 0.2 and 0.5 which shows that the nature of the oscillations of Chua's circuit at $\varepsilon$= 1000  (where $\lambda_m \approx$ 0) is regular. On the other hand the dynamics is chaotic when $p-q$ phase portrait is diffusive. $D(n)$ grows linearly with $n$, and the slope quantifies the nature of trajectories as chaotic for $\Phi$ = 1.0 and $\Phi$ = 1.5 in addition to $K$ being close to 1. Clearly at a fixed coupling ($\varepsilon$=1000) the dynamics of Chua's circuit go from regular when the time-scale of forcing is slow to chaotic when the time scales of drive is comparable to the response system. To ascertain the regular behavior of Chua's circuit in $\lambda_m \approx$ 0 region, we further analyze the trajectories by using other measures such as FFT and recurrence plots of the time-series data.

\subsection{FFT}
\begin{figure}[h]
	\centering
	\includegraphics[width=0.50\textwidth]{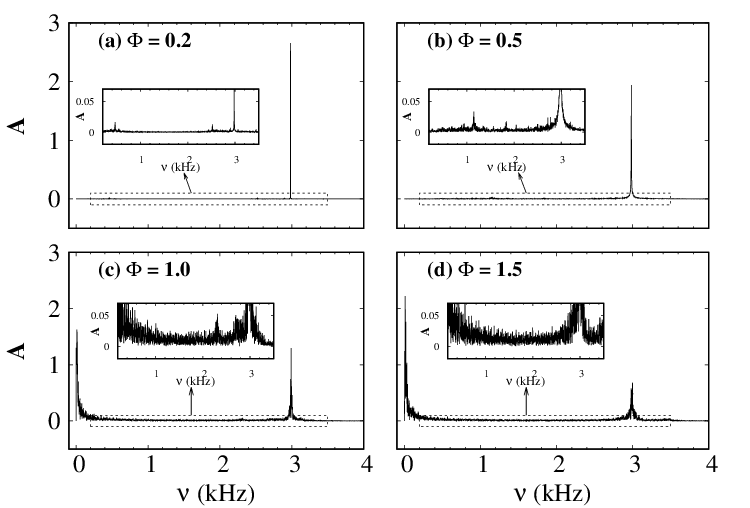}
	\label{fig:enter-label}
	\caption{FFT of the time series $V_1$ for different values of the $\Phi$ at $\varepsilon$ = 1000. The y-axis represents the amplitude, $A$ of the frequency components while the $x$-axis denotes the frequency, $\nu$. The inset shows magnified image of the amplitudes in the marked (box with dotted lines) frequency range.}    
	\label{fig:Fig17}   
\end{figure}

We do the fast Fourier transform (FFT) of the time-series $V_1$ to know the dominant frequencies. In Fig.~\ref{fig:Fig17} the FFT plots of the time series shows a clear single peak for $\Phi$ = 0.2 and 0.5 indicating periodic nature of the time series. As $\Phi$ increases other harmonics emerge and there is a spread in the FFT plot. The amplitude of these harmonics increase with increase in $\Phi$ and the dynamics becomes chaotic for higher values of $\Phi$ ( Fig.~\ref{fig:Fig17}(c)-(d)).

\subsection{Recurrence plots}

Recurrence plots (RPs)~\cite{eckmann1987} are used to study the recurrences of the phase space trajectories. Consider a dynamical system in \textit{d}-dimensional phase space represented by the trajectory \textit{$\{\bm{x}_i\}$} for \textit{i =} 1,...,N. The binary recurrence matrix is defined as:
\begin{equation}\label{eq:recur}
	R_{ij} = \Theta(\delta - \|\bm{x}_i - \bm{x}_j\| ), \qquad  i,j = 1,...,N,
\end{equation}
where $\Theta$ is the Heaviside function and $\left\| .\right\|$ is the norm of distance between two points. $\delta$ defines the threshold of the norm below which the pairs of phase space points will be considered to recur.

    \begin{figure}[h]
    	\centering
    	\includegraphics[width=0.50\textwidth]{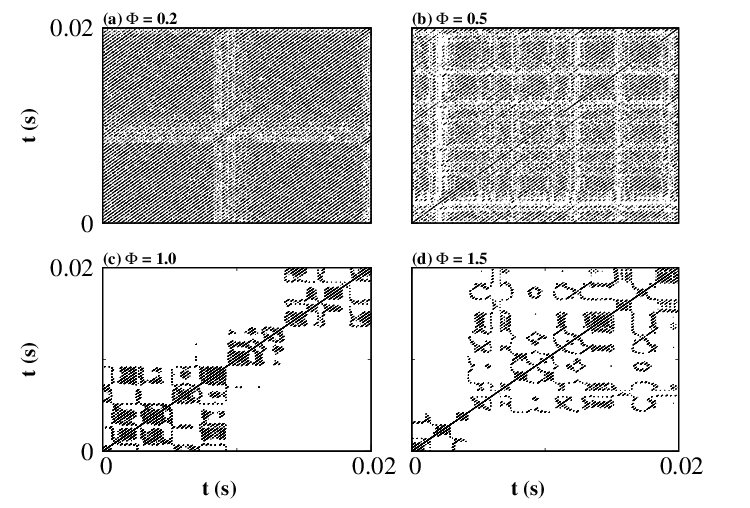}
    	\label{fig:enter-label}
    	\caption{RPs computed from the time-series of response Chua's circuit for different values of $\Phi$. The value of $\delta$ = 0.05. }    
    	\label{fig:Fig18}   
    \end{figure}

\begin{figure}[h]
	\centering
	\includegraphics[width=0.50\textwidth]{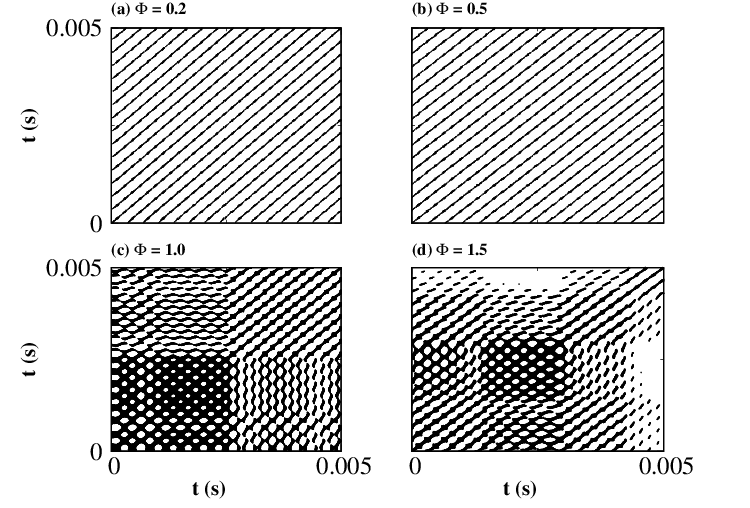}
	\label{fig:enter-label}
	\caption{The RP of the laminar phase of the trajectory of the driven Chua's circuit obtained for different values of $\Phi$ are shown. The $\delta$ value is taken to be 0.5 which is 10 \% of the diameter of the single scroll attractor.}    
	\label{fig:Fig19}   
\end{figure}

For our case we construct the RPs from the time-series of all three variables of the driven Chua's circuit for different values of $\Phi$ fixing $\varepsilon$ = 1000. The plots are shown in Fig.~\ref{fig:Fig18}. For slow drive ($\Phi$ = 0.2), we observe parallel equispaced lines in most region of RPs (see Fig.~\ref{fig:Fig18}(a)-(b)) which suggest periodic nature of the driven Chua's oscillator with intermittent bursts of irregular behavior when the system is evolved for long time. As the time-scale of the driving signal increases and approaches the time-scale of the response, we notice that the irregular behavior becomes more frequent and the trajectory becomes chaotic \cite{eckmann1987,kabiraj2012}. These RPs show signatures of intermittency route to chaos~\cite{pomeau1980} in the Chua's circuit dynamics as $\Phi$ increases. In order to investigate the type of intermittency in the response system in the coupling region where $\lambda_m$ = 0, we look at the behavior for periodic or laminar phase of the trajectory at different $\Phi$ as shown in Fig.~\ref{fig:Fig19}. For high frequency of the drive ($\Phi$ = 1.0, 1.5) we observe the laminar phase has black squares with perforated edges which suggest type-II intermittency \cite{intermittency2009}, while for low frequency drive ($\Phi$ = 0.2, 0.5) the periodic nature of the driven system can be seen. 

From the analysis of time-series of the response variables at a fixed coupling strength, we can conclude that a slow drive induces periodicity in the dynamics of the response system reflected in the value of largest CLE close to zero. As the time-scale of the drive increases and becomes comparable to that of the response, the response dynamics becomes irregular and chaotic. However, we would like to point out that our analysis do not exclude the possibility of quasi-periodicity in the response dynamics which needs to be investigated in more detail.

\section{Summary}  \label{sec:summary}

In this work we study the effect of variable time-scale of the drive on the response of Chua's circuit in chaotic regime. The driving signal is chaotic and drawn from a common base Colpitts oscillator. We observe that when the time-scale of drive is much slower than the response and the driving strength is above a certain threshold, the Chua's oscillator exhibits bistability with coexisting chaotic attractors. These multistable attractors are different from the double-scroll chaotic attractor of the uncoupled Chua's circuit in the sense that they are single scroll and occupy much smaller region in phase space. When an auxiliary system is considered to explore the prospects of generalized synchronization in such systems, we observe that the drive and auxiliary units show \emph {remote} synchronization where the two units have equal average frequency yet their dynamical variables are correlated to each other in different ways.  In addition to the scenario when the response and auxiliary systems are in complete synchrony (RS-I), there could be lag synchronization (RS-III) or correlated trajectories (RS-II and RS-IV) of the two Chua's circuits indicating new forms of remote synchrony between the uncoupled response and auxiliary units. The existence of complete synchrony between response and auxiliary implies generalised synchrony in these drive-response systems. The parameter region where RS-III and RS-IV coexist is characterized by largest conditional Lyapunov exponent close to zero. We analyzed this region using various tools such as 0-1 chaos test, FFT and recurrence plots and found that the dynamics of the Chua's circuit is periodic in the region of slow forcing and it becomes chaotic through intermittency as the time scale of the drive increases and become equivalent to that of response. The region where lag synchronization and correlated trajectories occur diminishes as the frequency of drive approaches the frequency of the response. Therefore, such kinds of states and hence RS are not possible if the drive and response variables evolve at similar time scales. We have also checked the case when the frequency of the drive is higher than the inherent frequency of the response and observed that in this set-up the dynamics of the response is dependent on the nature of the drive, for example a symmetric drive can give bistability with two coexisting single-scroll attractors whereas an asymmetric drive can favour one type of attractor over other. Although the results presented in this paper are for the dimensional model, the results have been found consistent with that obtained from the equivalent non-dimensional model. 

The multistabilty induced in the present system is the result of time-scale mismatch between drive and response, and hence the mechanism leading to this multistability seems to be different from the multistability induced by coupling reported earlier~\cite{ujjwal2016} where the coupling modifies the effective parameter to the values where the response system is inherently multistable. To check the generality of our results, we checked our results for different types of driving signals, for instance Chua's circuit driven by R$\ddot{o}$ssler oscillator and observed qualitatively similar states. It will be interesting to investigate the dynamics of Chua's circuit driven by periodic and quasiperiodic signal with variable frequency which is part of our ongoing research~\cite{ongoing}. Our preliminary investigation suggests that in addition to the multistability observed in the case of chaotic drive, we can also observe other kinds of multistable behavior involving resonance response of Chua's circuit. We believe that our results open up several interesting directions to explore, for instance the implications of reported multistable RS states in networks with hub structures. In a recent work by Gomes et al \cite{gomes2023} chaotic hysteresis is reported in Chua's circuit driven by slow voltage forcing. Hysteresis is often observed in the system that have multistability, so another possible direction to explore could be to look for the possibility of chaotic hysteresis in coupled systems with mismatched time scales.

\begin{acknowledgments}
TM and AS acknowledge the resources provided by PARAM Shivay computational facility at the Indian Institute of Technology, Varanasi. SRU is thankful to Banaras Hindu University for financial support under Institute of Eminence (IoE) scheme. We thank Awadhesh Prasad for useful discussions. 
\end{acknowledgments}



\begin{thebibliography}{28}
\bibitem{sim-details} For the simulations the initial conditions for Chua's and Colpitts oscillators are taken randomly in the range $-1.0$  to $1.0$ for $V_1$ and $V_2$ (in the unit of Volt) and $-0.005$ to $0.005$ for $I_L$ (in the unit of Ampere).
\bibitem{icfig3} The initial condition for the Chua's oscillator is taken to be ($V_1$, $V_2$, $I_L$) = (0.5,0.3,-0.001).
\end{thebibliography}

\end{document}